\documentclass[doublespacing]{elsart}

\usepackage{icarus}

\usepackage{natbib}

\bibpunct{(}{)}{;}{a}{,}{,}

\usepackage{graphicx}

\usepackage{amsmath}
\usepackage{commath}

\newcommand{\BibTeX}{ \textrm{B\kern-.05em\textsc{i\kern-.025em b}\kern-.08em
    T\kern-.1667em\lower.7ex\hbox{E}\kern-.125emX} }

\newcommand*\icarus{Icarus}
\newcommand*\jgr{J Geophys Res}
\newcommand*\grl{Geophys Res Lett}
\newcommand*\planss{Planetary Space Sci}
\newcommand*\jqsrt{J Quant Spec Rad Trans}
\newcommand*\apj{Astrophys J}
\newcommand*\apjs{Astrophys J Supp}
\newcommand*\aap{Astron Astrophys}
\newcommand*\mnras{Mon Not Royal Astro Soc}

\begin{document}

\begin{frontmatter}



\title{A Simple Model for Radiative and Convective Fluxes in Planetary Atmospheres}


\author[jpt]{Juan P. Tolento} and 
\author[tdr]{Tyler D. Robinson}

\address[jpt]{California Polytechnic State University, 
				San Luis Obispo, CA 93407 (U.S.A.)}
\address[tdr]{Northern Arizona University,
				Flagstaff, AZ 86005 (U.S.A.)}



%
%
%
%
%


\end{frontmatter}



\begin{flushleft}
\vspace{1cm}
Number of pages: \pageref{lastpage} \\
Number of tables: \ref{lasttable}\\
Number of figures: \ref{lastfig}\\
\end{flushleft}


\begin{pagetwo}{Planetary Radiative and Convective Fluxes}

Tyler D. Robinson \\
Northern Arizona University\\
Department of Physics and Astronomy\\
Box 6010\\
Flagstaff, AZ 86011-6010, USA. \\
\\
Email: tyler.robinson@nau.edu\\
Phone: 928-523-0350 \\
Fax: 928-523-1371

\end{pagetwo}

\begin{abstract}

One-dimensional (vertical) models of planetary atmospheres typically balance 
the net solar and internal energy fluxes against the net thermal radiative and 
convective heat fluxes to determine an equilibrium thermal structure.  Thus, simple 
models of shortwave and longwave radiative transport can provide insight into 
key processes operating within planetary atmospheres.  Here, we develop a simple, 
analytic expression for both the downwelling thermal and net thermal radiative fluxes 
in a planetary troposphere.  We assume that the atmosphere is non-scattering at 
thermal wavelengths and that opacities are grey at these same wavelengths.  
Additionally, we adopt an atmospheric thermal structure that follows a modified 
dry adiabat as well as a physically-motivated power-law relationship between grey 
thermal optical depth and atmospheric pressure.  To verify the accuracy of our 
analytic treatment, we compare our model to more sophisticated ``full physics'' tools 
as applied to Venus, Earth, and a cloudfree Jupiter, thereby exploring a diversity 
of atmospheric conditions conditions.  Next, we seek to better understand our analytic 
model by exploring how thermal radiative flux profiles respond to variations in key 
physical parameters, such as the total grey thermal optical depth of the atmosphere.  
Using energy balance arguments, we derive convective flux profiles for the tropospheres 
of all Solar System worlds with thick atmospheres, and propose a scaling  
that enables inter-comparison of these profiles.  Lastly, we use our analytic treatment 
to discuss the validity of other simple models of convective fluxes in planetary 
atmospheres.  Our new expressions build on decades of analytic modeling exercises in 
planetary atmospheres, and further prove the utility of simple, generalized tools 
in comparative planetology studies.

\end{abstract}




%
\section{Introduction}
%

The thermal structure of a planetary atmosphere is determined via complex 
energy and mass exchanges in radiative, advective, and diffusive processes.  
One-dimensional (vertical) models of planetary atmospheres seek to explain how 
radiative processes and the vertical convective transport of heat and condensible 
species combine to establish the average atmospheric structure of a world.  In 
these one-dimensional planetary climate models, treatments of radiative transport vary in 
complexity.  The most sophisticated radiation tools operate at high spectral resolution 
\citep{wordsworthetal2017} with full physics treatments of scattering processes 
\citep{robinson&crisp2018}, or may adopt lower-resolution correlated-$k$ techniques 
\citep{goodyetal1989,lacis&oinas1991}.  Alternatively, the simplest models may use a 
semi-grey two-stream approach \citep[e.g.,][]{mckayetal99}.  
Similarly, sophisticated treatments of convective transport can range from  
applications of planetary boundary layer physics \citep{mellor&yamada1974} to 
mixing length models \citep[e.g.,][]{gierasch&goody1968}.  Less-complex tools 
apply the relatively straightforward ``convective adjustment'' approach 
\citep{manabe&strickler1964}.  Many of these topics are discussed in a recent 
review by \citet{marley&robinson2015}.

Semi-grey, windowed-grey, and banded-grey radiative transfer techniques have a 
long history of application to planetary atmospheres 
\citep{robinson2015}.  Semi-grey or windowed-grey radiative 
transfer models have been used to explore the climate of modern and ancient Earth 
\citep{hart78,weaver&ramanathan1995,friersonetal2006,pelkowskietal2008}.  Similar semi-grey 
radiative approaches have been applied to studies of both the one-dimensional 
and three-dimensional structure of Titan's atmosphere \citep{mckayetal99,mitchelletal2006}, 
as well as to runaway greenhouse worlds \citep{nakajimaetal1992}.

Beyond the Solar System, the limited atmospheric data available combined with the 
explosion of interest in exoplanets has driven the need for simple parameterized 
\citep{madhusudhan&seager2009} or physically-based \citep{hubenyetal2003,hansen2008} 
atmospheric thermal structure models.  Building on the semi-grey radiative equilibrium 
solution applied to hot Jupiter exoplanets in \citet{guillot2010}, 
\citet{parmentier&guillot2014} developed a banded-grey ``picket fence'' 
radiative equilibrium model that can better reproduce the structure of hot Jupiter 
stratospheres \citep{parmentieretal2015}, where semi-grey radiative equilibrium models 
tend to overestimate atmospheric temperatures 
\citep[see explanation in][their Section 4.7]{pierrehumbert2011}.  
\citet{hengetal2012} extensively investigated the application of semi-grey radiative 
equilibrium models to hot Jupiter atmospheres, and these concepts have subsequently 
been adapted to include scattering \citep{hengetal2014} and ``picket fence'' 
thermal opacities \citep{mohandasetal2018}.

Markedly fewer studies have combined grey radiative transfer techniques with 
treatments of convection.  This may seem striking as convection is known to 
be critical to driving vertical transport, and can be related to effective 
upwelling windspeeds in planetary tropospheres \citep{gierasch&conrath1985}.
In a pair of early examples, \citet{sagan1969} and \citet{weaver&ramanathan1995} 
investigated the onset of convective instability in semi-grey and windowed-grey 
atmospheres.  

More recently, \citet{ozawa&ohmura1997} \citep[and also][]{wu&liu2010,herbertetal2013} used 
a semi-grey radiative transfer model to derive convective fluxes for Earth-like conditions 
under the ``maximum entropy production'' principle, wherein convective energy transport is 
postulated to maximize the local entropy production rate.  Indeed, \citet{ozawa&ohmura1997} 
showed that temperature profiles computed using the maximum entropy production 
principle were less steep and had lower surface temperatures than pure 
radiative equilibrium solutions, thus reproducing behaviors seen in models that 
adopt other treatments of convection \citep[e.g.,][]{manabe&strickler1964}. 
\citet{lorenz&mckay2003} used semi-grey radiative transport expressions to 
heuristically arrive at an analytic expression for the convective flux at a 
planetary surface, and showed that this expression could reproduce the 
convective fluxes computed by more complex models.  Finally, 
\citet{robinson&catling2012} produced analytic expressions for the 
thermal structure of a planetary atmosphere with a semi-grey radiative stratosphere 
overlying a convective troposphere whose structure follows a modified dry adiabat.  
These authors used this tool to understand the physics behind a common 0.1~bar 
tropopause pressure seen throughout the Solar System \citep{robinson&catling2014}.

In what follows, we extend the \citet{robinson&catling2012} model to include 
an analytic treatment of downwelling thermal radiative fluxes in a semi-grey 
atmosphere that includes a convective troposphere, and also introduce more 
realistic lower boundary conditions to a previous solution for the upwelling 
thermal radiative flux.  Taken together, these expressions for the upwelling 
and downwelling thermal radiative fluxes yield straightforward, physically-based 
expressions for both the net thermal flux and the convective heat flux in 
planetary tropospheres.  We validate our net thermal flux treatment against 
more sophisticated climate and radiative transfer models, and we apply our 
derived convective heat flux profiles to compare Solar System worlds.  Finally, 
we use our new model to comment on the maximum entropy production approach 
explored by \citet{ozawa&ohmura1997} and on the heuristic convective flux 
expression given by \citet{lorenz&mckay2003}.

%
\section{Theory}
%

In a one-dimensional, plane-parallel atmosphere, the grey two-stream Schwarzschild 
equations for non-scattering thermal radiative transport are \citep[][p.~84]{andrews2010},
\begin{equation}
    \frac{dF^{+}}{d\tau} = D\left( F^{+} - \sigma T^{4} \right) \ ,
\label{eqn:rteup}
\end{equation}
\begin{equation}
    \frac{dF^{-}}{d\tau} = -D\left( F^{-} - \sigma T^{4} \right) \ ,
\label{eqn:rtedn}
\end{equation}
where $\tau$ is the vertical grey thermal optical depth, $T$ is the temperature, 
$\sigma$ is the Stefan-Boltzmann constant ($5.67\times10^{-8}$~W~m$^{-2}$~K$^{-4}$), 
$D$ is the so-called diffusivity factor, and $F^{+}$ and $F^{-}$ are the upwelling 
and downwelling thermal radiative fluxes, respectively.  Recall that the diffusivity 
factor accounts for the integration of radiance over a hemisphere, and values 
spanning 1.5--2 are commonly adopted in the literature 
\citep{rodgers&walshaw1966,armstrong1968}.  Thus, given a temperature 
profile, $T\left(\tau\right)$, Equations~\ref{eqn:rteup} and \ref{eqn:rtedn} can 
be evaluated to yield the upwelling and downwelling thermal radiative flux profiles 
for an atmosphere.  While more sophisticated treatments of closure used to derive the 
two-stream equations have been detailed \citep{hengetal2014}, the extremely simple 
semi-grey approximation adopted in the present work argues against the need for moving  
to sophisticated two-stream models.

In the convective portion of a non-condensing atmosphere, the thermal structure 
follows Poisson's adiabatic state equation 
\citep[][their Equation~1.32]{catling&kasting2017},
\begin{equation}
    T = T_0 \left(\frac{p}{p_0}\right)^{(\gamma-1)/\gamma} \ ,
\end{equation}
where $T_0$ is a reference temperature at $p_0$ (taken to be, e.g., the surface, 
or the 1 bar pressure level in the atmosphere of a gaseous world), and $\gamma$ 
is the ratio of specific heats.  Following \citet{sagan1962}, we modify the dry 
adiabat to account for latent heat release or non-constant specific heats by 
introducing a parameter, $\alpha$, into the adiabatic state equation, giving, 
\begin{equation}
    T = T_0 \left(\frac{p}{p_0}\right)^{\alpha(\gamma-1)/\gamma} \ .
\label{eqn:adiabat}
\end{equation}
In the tropospheres of Solar System worlds above roughly the 1 bar pressure level, 
$\alpha$ is always of order unity, varying between 0.6 for Earth and 0.94 for 
Saturn.

To determine the upwelling and downwelling thermal radiative flux profiles in 
the convective portion of a planetary atmosphere, we wish to insert our adiabatic 
state equation into the radiative transfer expressions (Equations~\ref{eqn:rteup} 
and \ref{eqn:rtedn}).  A difficulty arises as the adiabatic equation is expressed 
with pressure (the natural vertical coordinate for planetary atmospheres) as the 
independent variable whereas the radiative flux expressions use grey thermal 
optical depth (the natural vertical coordinate for radiative transfer) as their 
independent variable.  Following \citet{pollack1969}, we relate pressure and grey 
thermal optical depth through a power law, with,
\begin{equation}
    \tau = \tau_0 \left( \frac{p}{p_0} \right)^{n} \ ,
\end{equation}
where $\tau_0$ is a reference grey thermal optical depth at $p_0$, and $n$ typically 
varies between 1 and 2, corresponding to opacities dominated by Doppler broadening 
versus either pressure broadening or collision-induced absorption, respectively.  
Although larger values of $n$ have been proposed for scenarios where thermal opacity 
sources condense out of the atmosphere 
\citep[e.g., water vapor in Earth's atmosphere;][]{weaver&ramanathan1995},  
models with $n=2$ have been shown to accurately reproduce spectrally-resolved models of 
thermal fluxes in Earth's atmosphere \citep{robinson&catling2014}.

Combining the adjusted adiabatic state equation with our power law relationship 
between pressure and grey thermal optical depth yields,
\begin{equation}
    T = T_0 \left(\frac{\tau}{\tau_0}\right)^{\alpha(\gamma-1)/\gamma n} =  T_0 \left(\frac{\tau}{\tau_0}\right)^{\beta/n}\ ,
\label{eqn:adiabat_tau}
\end{equation}
where we have defined $\beta = \alpha(\gamma-1)/\gamma$.  Note that, with this 
definition,
\begin{equation}
    \frac{d\ln T}{d\ln \tau} = \frac{\beta}{n} \ ,
\end{equation}
indicating that $\beta/n$ controls the 
steepness of the $T$-$\tau$ relation.  Inserting this into Equations~\ref{eqn:rteup} 
and \ref{eqn:rtedn}, and adopting an integrating factor of the form $e^{-D\tau}$ 
for the upwelling thermal flux and $e^{D\tau}$ for the downwelling thermal flux 
enables us to write the integral form of the solutions to Equations~\ref{eqn:rteup} 
and \ref{eqn:rtedn} as,
\begin{equation}
    \int d\left(F^{+}e^{-D\tau'} \right) = - D\sigma T_{0}^{4}\int \left( \frac{\tau}{\tau_0} \right)^{4\beta/n} e^{-D\left(\tau' - \tau \right)} d\tau' \,
\end{equation}
\begin{equation}
    \int d\left(F^{-}e^{D\tau'} \right) = D\sigma T_{0}^{4}\int \left( \frac{\tau}{\tau_0} \right)^{4\beta/n} e^{D\left(\tau' - \tau \right)} d\tau' \ .
\end{equation}
We must specify a set of boundary conditions to further solve these relations.  For the 
upwelling flux, we adopt a lower boundary condition of,
\begin{equation}
    F^{+}\left( \tau_0 \right) = (1 + f) \sigma T_{0}^{4} ,
\end{equation}
with,
\begin{equation}
f = 
  \begin{cases}
    0, & \text{solid lower boundary}  \\ 
    \frac{4\beta}{n} \frac{1}{D\tau_0}, & \text{diffuse lower boundary} 
  \end{cases}
\ ,
\end{equation}
where the diffuse lower boundary scenario enables thermal flux from 
layers below $\tau_0$ to contribute to $F^{+}\left( \tau_0 \right)$ and 
also ensures that the radiation diffusion limit is obeyed in opaque 
conditions.  For the downwelling flux, at the top of the convective zone 
(i.e., at the so-called radiative-convective boundary, only below which 
does Equation~\ref{eqn:adiabat} apply), located at $\tau_{\rm{rc}}$, we 
assume that the downwelling thermal flux is from an overlying radiative 
portion of the atmosphere, which we take as 
$F^{-}\left(\tau_{\rm{rc}}\right)$.

Inserting our boundary conditions, the integral solutions for the upwelling 
and downwelling thermal fluxes become, 
\begin{equation}
    \int_{(1+f)\sigma T_{0}^{4}e^{-D\tau_0}}^{F^{+}(\tau)e^{-D\tau}} d\left(F^{+}e^{-D\tau'} \right) = - D\sigma T_{0}^{4}\int_{\tau_0}^{\tau} \left( \frac{\tau}{\tau_0} \right)^{4\beta/n} e^{-D\left(\tau' - \tau \right)} d\tau' \,
\end{equation}
\begin{equation}
    \int_{F^{-}(\tau_{\rm{rc}})e^{-D\tau_{\rm{rc}}}}^{F^{-}(\tau)e^{-D\tau}} d\left(F^{-}e^{D\tau'} \right) = D\sigma T_{0}^{4}\int_{\tau_{\rm{rc}}}^{\tau} \left( \frac{\tau}{\tau_0} \right)^{4\beta/n} e^{D\left(\tau' - \tau \right)} d\tau' \ .
\end{equation}
Or, after simplification,
\begin{equation}
    F^+\left( \tau \right) = (1+f)\sigma T_0^4 e^{-D\left(\tau_0 - \tau\right)} + D\sigma T_0^4 \int_{\tau}^{\tau_0} \left( \frac{\tau'}{\tau_0} \right)^{4\beta/n} e^{-D\left(\tau' - \tau \right)} d\tau' \ ,
\label{eqn:Fupint}
\end{equation}
\begin{equation}
    F^-\left( \tau \right) = F^{-}(\tau_{\rm{rc}}) e^{-D\left(\tau - \tau_{\rm{rc}}\right)} + D\sigma T_0^4 \int_{\tau_{\rm{rc}}}^{\tau} \left( \frac{\tau'}{\tau_0} \right)^{4\beta/n} e^{-D\left(\tau - \tau' \right)} d\tau' \ ,
\label{eqn:Fdnint}
\end{equation}
where, for $F^+$, the first term on the right hand side represents thermal flux 
that is exponentially attenuated away from the lower boundary and, for $F^-$, the 
first term on the right hand side represents thermal flux that is exponentially 
attenuated away from the upper boundary (i.e., the exponential attenuation of 
thermal flux from the overlying radiative portion of the atmosphere).  For both 
expressions, the second term on the right hand side represents thermal flux emitted 
from an atmospheric layer at $\tau'$ and exponentially attenuated to a layer at $\tau$.

By solving the integral in Equation~\ref{eqn:Fupint}, \citet{robinson&catling2012} 
showed that the analytic solution for the upwelling thermal radiative flux is,
\begin{equation}
\begin{split}
    F^+\left( \tau \right) = & ~(1+f)\sigma T_0^4 e^{-D\left(\tau_0 - \tau\right)} ~~ + \\ 
                             & ~\frac{\sigma T_0^4 e^{D\tau}}{\left(D \tau_0 \right)^{4\beta/n}}\left[ \Gamma\left(1+4\beta/n,D\tau\right) - \Gamma\left(1+4\beta/n,D\tau_0\right) \right] ,
\end{split}
\label{eqn:Fup}
\end{equation}
where $\Gamma(a,x)$ is the upper incomplete gamma function, and we have now included 
a boundary condition that allows for the treatment of a diffuse lower boundary.  
These authors, however, did not explore analytic solutions for the downwelling 
thermal radiative flux, which we now provide.  Inspecting Equation~\ref{eqn:Fdnint}, 
we seek a solution to integrals of the form,
\begin{equation}
    D\int_{\tau_{\rm{rc}}}^{\tau} \left( \tau' \right)^{4\beta/n} e^{D\tau'}  d\tau' \ .
\label{eqn:integral}
\end{equation}
By substituting $u$ for $-D\tau'$ in the integrand above, the corresponding indefinite 
integral takes the form of an upper incomplete gamma function.  Thus, the solution to the 
integral in Equation~\ref{eqn:integral} is the difference of two upper incomplete gamma 
functions,
\begin{equation}
  \frac{e^{-i\pi 4\beta/n}}{D^{4\beta/n}}\left[ \Gamma(1+4\beta/n, -D\tau)-\Gamma(1+4\beta/n, -D\tau_{\rm{rc}}) \right] \ .
\label{eqn:usub}
\end{equation}
While evaluations of the upper incomplete gamma function with negative arguments produce 
complex numbers, we know that the integral in Equation~\ref{eqn:Fdnint} must yield real 
numbers for the downwelling thermal flux.  Combining Equations~\ref{eqn:Fdnint} and~\ref{eqn:usub}, 
while taking the magnitude of the latter, gives an analytic expression for the downwelling 
thermal radiative flux as, 
\begin{equation}
\begin{split}
    F^-\left( \tau \right) = & ~F^{-}(\tau_{\rm{rc}}) e^{-D\left(\tau - \tau_{\rm{rc}}\right)} ~~ + \\ 
                             & ~\frac{\sigma T_0^4 e^{-D\tau}}{\left(D \tau_0 \right)^{4\beta/n}}\abs{ \Gamma\left(1+4\beta/n,-D\tau\right) - \Gamma\left(1+4\beta/n,-D\tau_{\rm{rc}}\right) } .
\end{split}
\label{eqn:Fdn}
\end{equation}

Combining our analytic expressions for the upwelling and downwelling thermal radiative 
fluxes allows us to express the net thermal radiative flux,
\begin{equation}
    F_{\rm{net}}\left( \tau \right) = F^+\left( \tau \right) - F^-\left( \tau \right) \ ,
\end{equation}
as,
\begin{equation}
\begin{split}
    F_{\rm{net}}\left( \tau \right) = & ~\frac{\sigma T_0^4}{\left(D \tau_0 \right)^{4\beta/n}} \Big[ (1+f)\left(D \tau_0 \right)^{4\beta/n}e^{-D( \tau_0 - \tau )} ~~ + \\
                                      & ~ \Gamma\left(1+4\beta/n,D\tau\right)e^{D\tau} - \Gamma\left(1+4\beta/n,D\tau_0\right)e^{D\tau} ~~ - \\ 
                                      & ~ \abs{ \Gamma\left(1+4\beta/n,-D\tau\right)e^{-D\tau} - \Gamma\left(1+4\beta/n,-D\tau_{\rm{rc}}\right)e^{-D\tau}} \Big] ~~ - \\
                                      & ~F^{-}(\tau_{\rm{rc}}) e^{-D\left(\tau - \tau_{\rm{rc}}\right)} \ .
\end{split}
\label{eqn:irfnet}
\end{equation}
Note that we have factored out $\sigma T_0^4/(D\tau_0)^{4\beta/n}$ from all terms except for the 
term which represents downwelling thermal radiative flux attenuated away from the radiative-convective 
boundary. Critically, if the net solar flux profile, $F_{\rm{net}}^{\odot}$, can be expressed or parameterized 
in terms of the grey thermal optical depth, then the atmosphere of a world with an internal energy flux, 
$F_{\rm{i}}$, is in equilibrium when,
\begin{equation}
    F_{\rm{net}}\left( \tau \right) + F_{\rm{c}}\left( \tau \right) = F_{\rm{net}}^{\odot}\left( \tau \right) + F_{\rm{i}} \ ,
\label{eqn:eqm}
\end{equation}
where $F_{\rm{c}}$ is the convective energy flux, and all fluxes have been taken as non-negative.

In the optically thick limit, the net thermal radiative flux should obey the radiation diffusion 
limit, with,
\begin{equation}
    F_{\rm{net}}\left( \tau \right) \approx \frac{8\beta}{n} \frac{\sigma T_0^4}{\left(D\tau_0\right)^{4\beta/n}} \cdot \left( D\tau \right)^{4\beta/n-1} \ .
\label{eqn:raddiff}
\end{equation}
Inspecting Equation~\ref{eqn:irfnet}, when $\tau_{\rm{rc}} \ll \tau \ll \tau_0$, we have,
\begin{equation}
\begin{split}
    F_{\rm{net}}\left( \tau \right) \approx & ~ \frac{\sigma T_0^4}{\left(D \tau_0 \right)^{4\beta/n}}\Big[ \Gamma\left(1+4\beta/n,D\tau\right)e^{D\tau} ~~ - \\
                                            & \abs{ \Gamma\left(1+4\beta/n,-D\tau\right)e^{-D\tau} }\Big]  \ .
\end{split}
\end{equation}
Thus, evidently, we have,
\begin{equation}
    \lim_{z\to\infty} \Big[ \Gamma\left(1+a,z\right)e^{z} - \abs{ \Gamma\left(1+a,-z\right)e^{-z} } \Big] = 2az^{a-1} \ ,
\end{equation}
which later numerical results will demonstrate.

%
\section{Validation}
\label{sec:valid}
%

The analytic expression for the net thermal radiative flux (Equation~\ref{eqn:irfnet}), while convenient, 
can only be shown to be useful through comparisons to observations or more sophisticated models.  Here, 
we compare our analytic treatment to net thermal radiative fluxes derived from one-dimensional 
spectrally-resolved (``full physics'') models of Venus, Earth, and a cloudfree Jupiter.  The Venus 
comparison tests our expression in very opaque conditions, while the Earth case tests applications 
in a relatively infrared-transparent atmosphere.  The cloudfree Jupiter comparison spans both 
regimes, and also explores a scenario with a diffuse lower boundary.

In all three scenarios below, the approach to deriving key parameters in Equation~\ref{eqn:irfnet} 
is the same.  The known atmospheric composition and thermal structure enables straightforward 
calculation of $\beta$ as well as $T_0$, and an appropriate value of $n$ is adopted.  The values of 
$\tau_0$, $\tau_{\rm{rc}}$, and $F^-\left(\tau_{\rm{rc}}\right)$ are obtained from application of 
the \citet{robinson&catling2012} analytic radiative-convective thermal structure model, which 
self-consistently solves for each of these three parameters.  To apply the 
\citet{robinson&catling2012} model, the net solar radiative flux profile must be parameterized as a 
sum of two exponentials, with,
\begin{equation}
    F_{\rm{net}}^{\odot}\left( \tau \right) = F^\odot_1e^{-k_1\tau} + F^{\odot}_2e^{-k_2\tau} \ ,
\label{eqn:fnetsol}
\end{equation}
where $k_i$ controls the strength of attenuation of 
solar flux, $F^\odot_i$, in one of the two shortwave channels.
We fit a function of this form to the net solar radiative flux profile computed by the 
spectrally-resolved models described below.  This approach is different from that of 
\citet{robinson&catling2014}, who adopted parameters in Equation~\ref{eqn:fnetsol} 
appropriate for dividing the net solar radiative flux into a stratospheric and tropospheric 
channel, and, for the former channel, selected a value for $k$ designed to reproduce 
the temperature at the stratopause.  In the validations below we do not distinguish between 
the solar radiative flux absorbed in the stratosphere versus troposphere, we merely seek 
well-fit reproductions of the net solar radiative flux profile using 
Equation~\ref{eqn:fnetsol}.

\subsection{Earth}

Our validation against Earth adopts a widely-used representative one-dimensional thermal 
structure profile for our planet \citep{mcclatcheyetal1972}.  To derive spectrally-resolved 
solar and thermal fluxes, we use the well-validated Spectral Mapping Atmospheric 
Radiative Transfer (SMART) model \citep[developed by D.~Crisp;][]{meadows&crisp1996}.  
We apply the SMART model to clearsky ocean, thick low-cloud, and thin high-cloud scenarios.  A 
weighted combination of the radiative flux profiles from these simulations (with 
25\% clearsky ocean, 35\% thick low-cloud, and 40\% thin high-cloud) yields a top-of-atmosphere 
net solar radiative flux of 240~W~m$^{-2}$ --- consistent with Earth's Bond albedo of 
0.3 and insolation of 1360~W~m$^{-2}$ --- and a top-of-atmosphere net thermal radiative flux that 
is in equilibrium with this absorbed solar flux.

We adopt $n=2$ for our analytic models of Earth.  While others have argued for a 
steeper $\tau$-$p$ relationship based on the decreasing mixing ratio of water vapor in 
Earth's troposphere \citep[e.g.,][who adopt $n=4$]{friersonetal2006}, \citet{robinson&catling2014} 
showed the $n=2$ offers the best reproduction of upwelling and downwelling thermal 
radiative fluxes for Earth's atmosphere.  Given this $\tau$-$p$ scaling, and the net 
solar radiative flux profile from the spectrally-resolved model, we find a best-fit of 
Equation~\ref{eqn:fnetsol} with $F^\odot_1=25$~W~m$^{-2}$, $F^\odot_2=220$~W~m$^{-2}$, $k_1=28$, and 
$k_2=0.13$.  Applying the \citet{robinson&catling2012} model yields yields $\tau_0=2.8$ and $\tau_{\rm{rc}}=0.16$, where we have adopted $p_0=1.013$~bar, $\gamma=1.4$, $T_0=294$~K, and $\alpha=0.6$, where the latter two parameters are designed to match the \citet{mcclatcheyetal1972} profile.

\begin{figure}[p!]
\begin{center}
\includegraphics[width=4in]{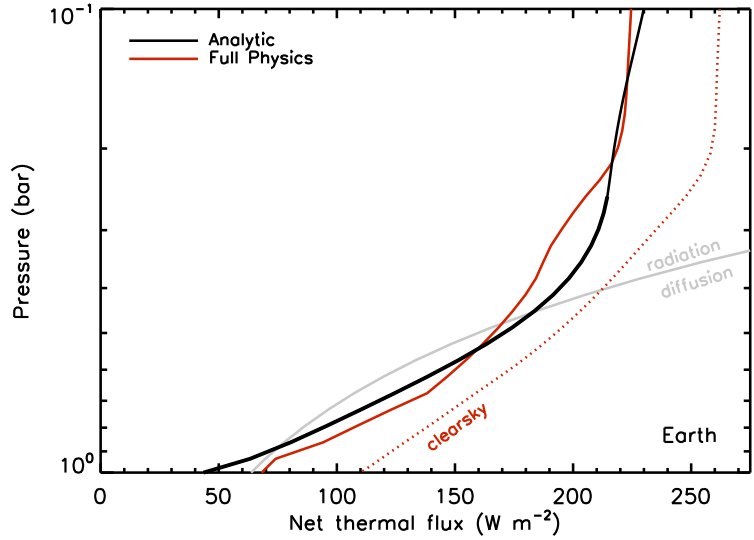}
\caption{
	\label{fig:earth_valid}
	Comparison between the analytic expression for net thermal radiative 
	flux (black; Equation~\ref{eqn:irfnet}) and the net thermal radiative flux 
	computed from a full physics radiative transfer model (red).  For the 
	full physics model, weighted partially-cloudy (solid) and clearsky (dotted) cases 
	are shown.  For the analytic treatment, the thickened portion of the 
	curve indicates the convective portion of the atmosphere.  Also shown 
	is the net thermal radiative flux computed in the radiative diffusion 
	limit (grey; Equation~\ref{eqn:raddiff}).
	}
\end{center}
\end{figure}

A comparison between the net thermal radiative fluxes computed by the full physics 
model versus our analytic treatment is shown in Figure~\ref{fig:earth_valid}.  For the 
full physics model, we include both the weighted partially cloudy scenario as well as 
a clearsky calculation.  Also shown is a net thermal radiative flux profile in the 
radiation diffusion limit (Equation~\ref{eqn:raddiff}).  Our analytic treatment 
reproduces both the shape and magnitude of the full physics partially cloudy model, 
whereas the shape of the radiation diffusion expression is a poor match.  This latter 
finding simply stems from the grey Earth model not being particularly opaque to thermal 
radiation, with $\tau_0=2.8$.

\subsection{Venus}

To compare against Venus, we adopt thermal structure and radiative flux profiles from 
a new one-dimensional full physics radiative-convective model \citep{robinson&crisp2018}.  
Critically, this model has been shown to reproduce Venus' observed thermal structure 
\citep{tellmannetal2009} as well as the Venus International Reference Atmosphere model 
\citep{moroz&zasova1997}.  Additionally, this full physics tool reproduces probe-derived 
observations of the net thermal \citep{revercombetal1985} and solar \citep{tomaskoetal1980} 
radiative flux profiles in Venus' atmosphere.

\citet{robinson&catling2012} proposed that the $\tau$-$p$ relationship in the deep atmosphere 
of Venus is likely to be less steep than $n=2$ due to strong overlap of absorption lines.  
Indeed, here we find that adopting $n=2$ yields a surface net thermal flux that is over 
an order of magnitude smaller than the value from our full physics model.  Thus, we explore 
a model with $n=1$.

\begin{figure}[p!]
\begin{center}
\includegraphics[width=4in]{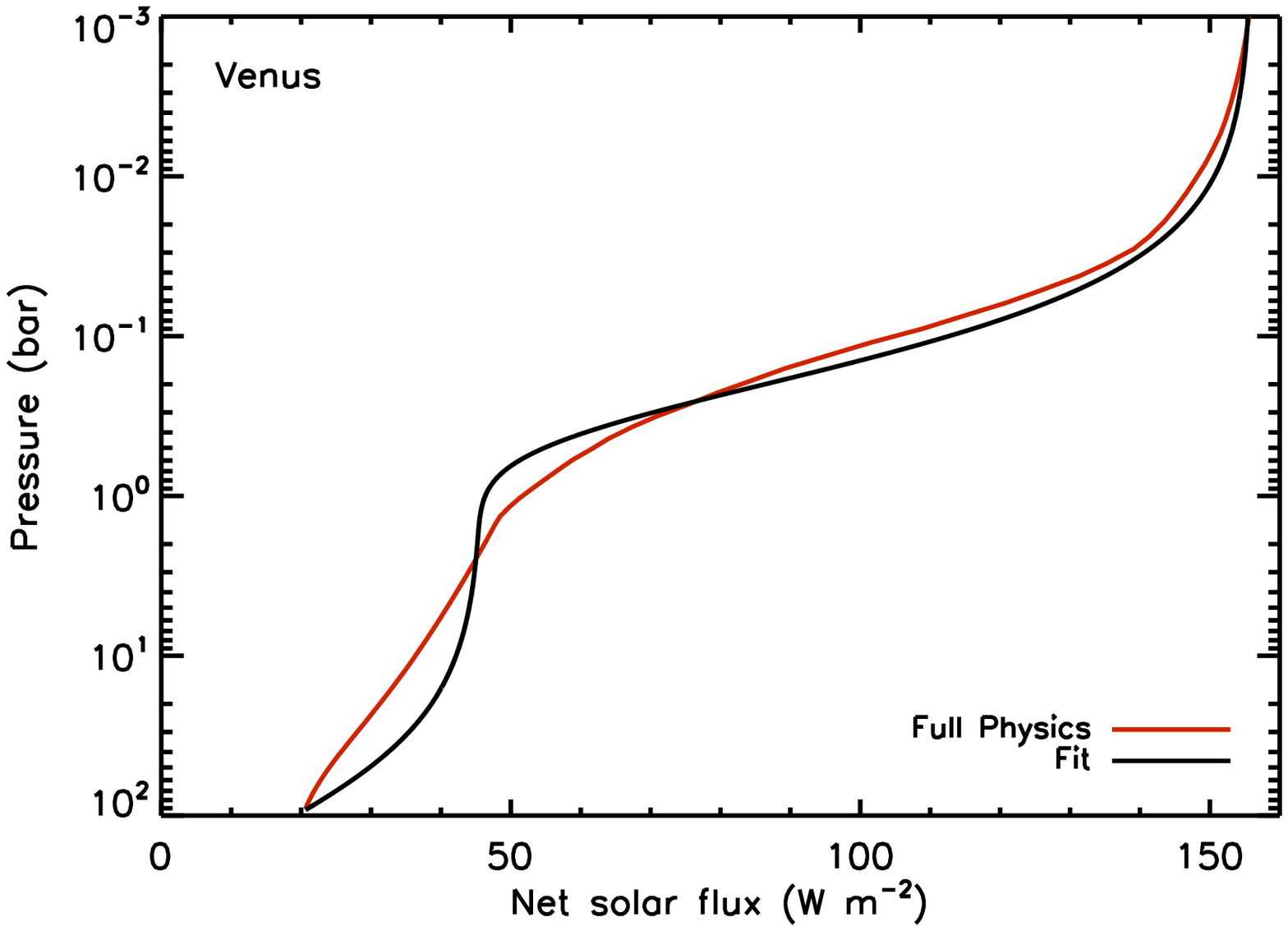}
\includegraphics[width=4in]{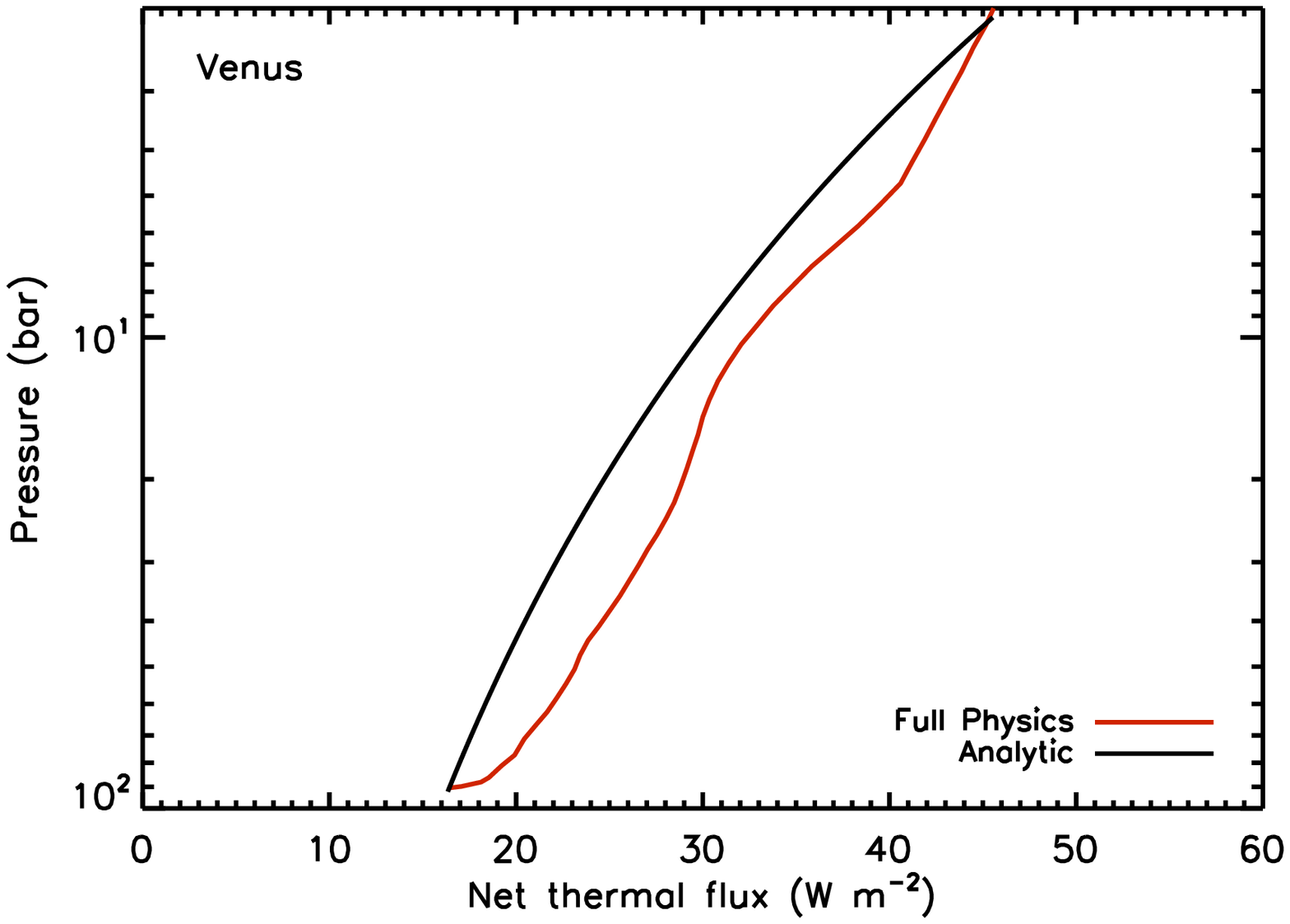}
\caption{
	\label{fig:venus_valid}
	Top panel shows our parameterized fit (black) to the full physics net solar flux 
	profile (red) for Venus using Equation~\ref{eqn:fnetsol}.  Bottom panel is the same 
	as Figure~\ref{fig:earth_valid} but for Venus, highlighting the atmosphere below 
	the Cytherean clouds.  The radiation diffusion limit is omitted as strongly opaque 
	atmospheric conditions imply our analytic expression (Equation~\ref{eqn:irfnet}) 
	completely overlaps this limit.
	}
\end{center}
\end{figure}

The net solar flux profile beneath the Cytherean clouds is only a weak function of 
pressure, decreasing by only about 50\% over two orders of magnitude in pressure.  While this 
profile is not well reproduced by an exponential function, Figure~\ref{fig:venus_valid} shows 
a reasonable reproduction of the net solar flux profile using Equation~\ref{eqn:fnetsol}.  
This reproduction adopts $F^\odot_1=110$~W~m$^{-2}$, $F^\odot_2 = 50$~W~m$^{-2}$, $k_1=0.5$, and 
$k_2=9\times10^{-4}$.  With our parameterized net solar flux profile and setting $p_0=92.1$~bar, 
$\gamma=1.3$, $\alpha=0.8$, and $T_0=730$~K, we find $\tau_{\rm{rc}}=20$ and $\tau_0=900$.

We compare our analytic models of the net thermal radiative flux to 
that computed by the full physics model in Figure~\ref{fig:venus_valid}.  The radiation 
diffusion limit is not shown as the large thermal grey optical depths cause 
Equation~\ref{eqn:irfnet} to completely overlap the radiation diffusion expression.  
Critically, our analytic model reproduces the shape of the full physics simulation, 
further justifying our choice of $n=1$.  Additionally, the analytic model reproduces 
the magnitude of the net thermal flux throughout the deep atmosphere below the Venusian 
clouds.

\subsection{Cloudfree Jupiter}

Finally, to explore a scenario with a diffuse lower boundary, we compare 
Equation~\ref{eqn:irfnet} to results from a widely-used one-dimensional climate 
model for gaseous Solar System planets and exoplanets 
\citep{marleyetal1999,fortneyetal2008}.  For this spectrally-resolved climate 
model, we adopt appropriate modern Jupiter parameters from \citet{fortneyetal2011}.  
We omit clouds from our model runs for two key reasons.  
First, and as discussed in \citet{fortneyetal2011}, modeled clouds lack a treatment 
for the absorptive chromophores in Jupiter's aerosols, thereby creating a simulated world 
with an unphysically large Bond albedo.  Second, implementing clouds within this 
full physics model introduces a large number of free parameters, many of which are 
poorly constrained.  As our primary goal is just to compare a simple analytic treatment 
with a more sophisticated numerical treatment of climate and radiation, we argue 
that clouds introduce unnecessary complexities.

\begin{figure}[p!]
\begin{center}
\includegraphics[width=4in]{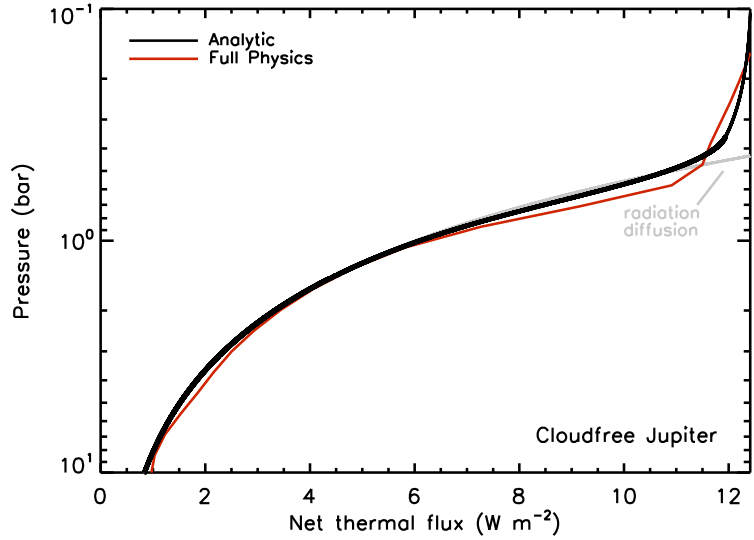}
\caption{
	\label{fig:jupiter_valid}
	Same as Figure~\ref{fig:earth_valid} but for a cloudfree Jupiter.
	}
\end{center}
\end{figure}

The convective portion of the cloudfree Jupiter atmosphere (where 
Equation~\ref{eqn:irfnet} applies) is likely to be in the regime where pressure-broadening 
and pressure-induced absorption (especially due to H$_2$-H$_2$ and H$_2$-He pairs) will 
dominate the opacity, implying $n=2$ is appropriate.  Adopting this value of $n$, the net 
solar radiative flux profile from our cloudfree Jupiter climate simulation 
is well reproduced with Equation~\ref{eqn:fnetsol}, yielding $F^\odot_1=3.4$~W~m$^{-2}$, 
$F^\odot_2=3.6$~W~m$^{-2}$, $k_1=0.28$, and $k_2=2.3\times10^{-3}$.  Using these within the 
context of the \citet{robinson&catling2012} model --- adopting $p_0=1.1$~bar, 
$\gamma=1.4$, $\alpha=1$, and $T_0=160$~K, all taken from the equilibrium climate 
solution from the full physics model --- yields $\tau_0=5.3$ and $\tau_{\rm{rc}}=0.58$.  
A comparison between the net thermal radiative fluxes from our analytic model versus 
the full physics model are shown in Figure~\ref{fig:jupiter_valid}.  The analytic model 
follows the radiation diffusion limit below about $0.5$~bar (where $D\tau$ is 
roughly 2), and the result from Equation~\ref{eqn:irfnet} is an excellent reproduction 
of the full physics model in both the convective portion of the atmosphere and the 
lower (radiative) portion of the stratosphere.

%
\section{Model Behavior}
%

\begin{figure}[p!]
\begin{center}
\includegraphics[width=4in]{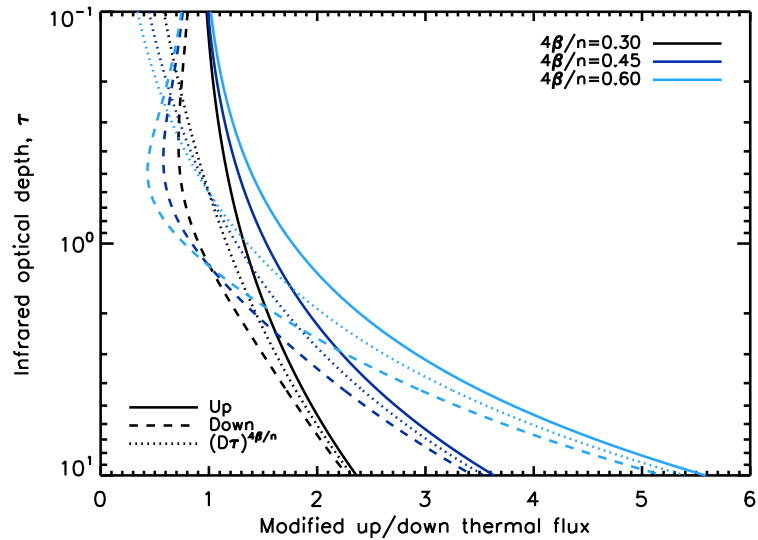}
\caption{
	\label{fig:demo_fup_fdn}
	Profiles of the ``modified'' upwelling (solid) and downwelling (dashed) 
	thermal radiative flux profiles, obtained by dividing 
	Equations~\ref{eqn:modFup} and \ref{eqn:modFdn} by 
	$\sigma T_0^4/\left(D\tau_0\right)^{4\beta/n}$.  Profiles are shown for 
	a range of physically-motivated values of $4\beta/n$.  Also shown are 
	power-laws of the form $\left(D\tau\right)^{4\beta/n}$ (dashed), demonstrating 
	that the modified upwelling and downwelling thermal radiative fluxes approach 
	$\sigma T\left(\tau\right)^4$ in optically thick conditions.
	}
\end{center}
\end{figure}

Intuition for the shape of the profiles for the upwelling, downwelling, and net thermal 
radiative fluxes can be obtained by manipulating Equations~\ref{eqn:Fup}, \ref{eqn:Fdn}, 
and \ref{eqn:irfnet}, and by exploring the resulting expressions for a range of 
physically-motivated values.  Inspecting Equations~\ref{eqn:Fup} and \ref{eqn:Fdn}, for 
$\tau_{\rm{rc}} \ll \tau \ll \tau_0$ we have,
\begin{equation}
    F^+\left( \tau \right) \approx \frac{\sigma T_0^4}{\left(D \tau_0 \right)^{4\beta/n}} \Gamma\left(1+4\beta/n,D\tau\right)e^{D\tau}  ,
\label{eqn:modFup}
\end{equation}
\begin{equation}
    F^-\left( \tau \right) \approx \frac{\sigma T_0^4}{\left(D \tau_0 \right)^{4\beta/n}} \abs{\Gamma\left(1+4\beta/n,-D\tau\right)}e^{-D\tau}  ,
\label{eqn:modFdn}
\end{equation}
indicating that curves of the form $\Gamma\left(1+4\beta/n,D\tau\right)e^{D\tau}$ for the 
upwelling case, and $\abs{\Gamma\left(1+4\beta/n,-D\tau\right)}e^{-D\tau}$ for the downwelling 
case, help indicate the shape of the thermal radiative flux profiles.  We refer to such functions 
as the ``modified'' upwelling or downwelling flux, and plot several such pairs of curves in 
Figure~\ref{fig:demo_fup_fdn} for a range of physically-motivated values of $4\beta/n$.  Recall 
that $\beta/n$ indicates the steepness of the $T$-$\tau$ relationship, so that $4\beta/n$ controls 
the gradient in the Stefan-Boltzmann emittance.

Regarding the net thermal radiative flux given in Equation~\ref{eqn:irfnet}, we can add 
$F^-\left(\tau_{\rm{rc}}\right)e^{-D\left(\tau-\tau_{\rm{rc}}\right)}$ from both sides and 
divide by $\sigma T_0^4$ to yield a ``modified'' net thermal 
radiative flux of the form,
\begin{equation}
\begin{split}
    \frac{F_{\rm{net}}\left( \tau \right) + F^{-}(\tau_{\rm{rc}}) e^{-D\left(\tau - \tau_{\rm{rc}}\right)}}{\sigma T_0^4} = ~~~~~~~~~~~~~~~ \\
    ~(1+f)e^{-D( \tau_0 - \tau )} ~~ + \\
    ~ \left(D \tau_0 \right)^{-4\beta/n} \Big[ \Gamma\left(1+4\beta/n,D\tau\right)e^{D\tau} - \Gamma\left(1+4\beta/n,D\tau_0\right)e^{D\tau} ~~ - \\ 
    \abs{ \Gamma\left(1+4\beta/n,-D\tau\right)e^{-D\tau} - \Gamma\left(1+4\beta/n,-D\tau_{\rm{rc}}\right)e^{-D\tau}} \Big] \ .
\end{split}
\label{eqn:modirfnet}
\end{equation}
Thus, the function on the right-hand side of this expression (which depends only on 
$4\beta/n$, $\tau_0$, $\tau_{\rm{rc}}$, and $\tau$) indicates the shape of the net thermal 
radiative flux profile.  Furthermore, this function will approach the true shape of the 
net thermal radiative flux profile for large $D\left(\tau - \tau_{\rm{rc}}\right)$, as 
the upper boundary condition flux term, 
$F^-\left(\tau_{\rm{rc}}\right)e^{-D\left(\tau-\tau_{\rm{rc}}\right)}$, is exponentially 
attenuated away from the radiative-convective boundary.

To understand the overall behavior of the net thermal radiative flux profile, we plot the function with nominal values ($\tau_0=5$, $\tau_{\rm{rc}}=.5$, and $4\beta/n=.45$) and compare them to  profiles where a single parameter is varied. We consider scenarios with either a solid or diffuse lower boundary condition. Typically, scenarios with different lower boundary conditions only diverge from one another when $\tau$ approaches $\tau_0$.

\begin{figure}[p!]
\begin{center}
\includegraphics[width=4in]{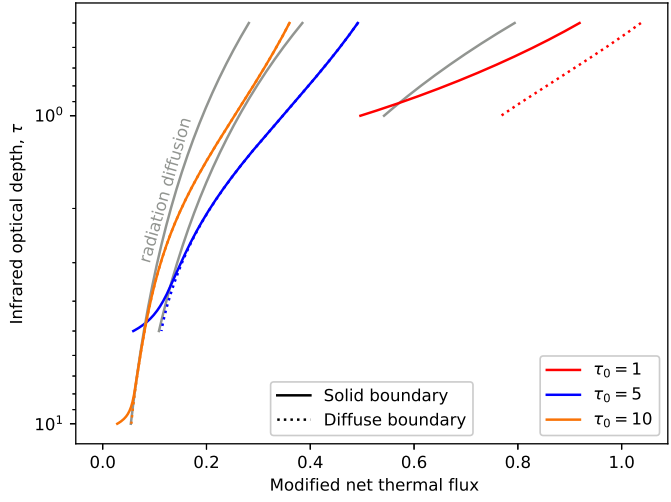}
\caption{
    \label{fig:varytau0}
    Comparison of the modified net thermal flux (Equation~\ref{eqn:modirfnet}) with varying values of $\tau_0$ and fixed $\tau_{\rm{rc}}=0.45$ and $4\beta/n=0.45$. Each value is explored using both the solid boundary and the diffuse boundary. Each profile is plotted from $\tau_{\rm{rc}}$ to their respective value of $\tau_0$. The x-axis represents our modified net thermal flux on a linear scale while the y-axis shows the infrared optical depth on a logarithmic scale. 
    }
\end{center}
\end{figure}

Modified net radiative flux profiles for cases where $\tau_0$ is varied to either $\tau_0 = 1$ or $\tau_0 = 10$ are shown in Figure~\ref{fig:varytau0}. As the radiation diffusion limit is dependent on $\tau_0$, each profile also has a unique limit,  that (for $\tau_0 = 5$ and $\tau_0 = 10$) meets with the modified net flux for the diffuse boundary at large values of $\tau$. The case where $\tau_0 = 1$ is not opaque enough for the atmosphere to reach the radiation diffusion limit. Furthermore, while the two larger values of $\tau_0$ have solid and diffuse profiles that match at small values of $\tau$, for $\tau_0=1$, the atmosphere is transparent enough that the lower boundary condition affects the entire modified net thermal flux profile.

\begin{figure}[p!]
\begin{center}
\includegraphics[width=4in]{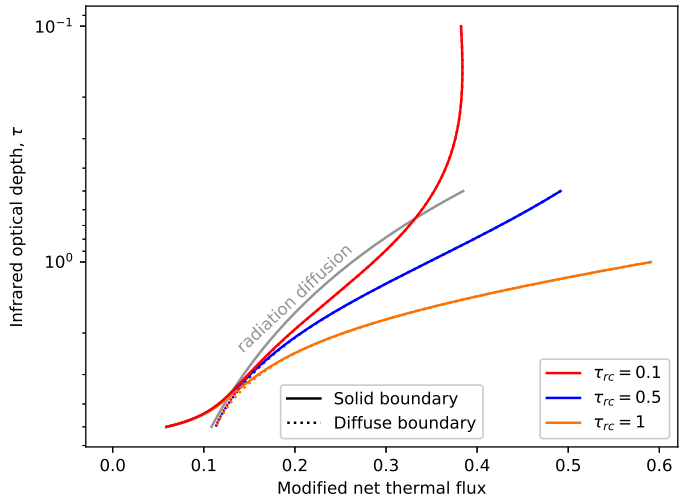}
\caption{
    \label{fig:varytaurc}
    Same as Figure~\ref{fig:varytau0}, except varying $\tau_{\rm{rc}}$.  We fix $\tau_0=5$ and $4\beta/n=0.45$.
    }
\end{center}
\end{figure}

After returning to our nominal value of $\tau_0 = 5$, we modify $\tau_{\rm{rc}}$ from $\tau_{\rm{rc}} = 0.5$ to values of $\tau_{\rm{rc}} = 0.1$ and  $\tau_{\rm{rc}} = 1$, the profiles of which are seen in Figure~\ref{fig:varytaurc}. In this case, all values of $\tau_{\rm{rc}}$ share the same radiative diffusion limit as it is not dependent on $\tau_{\rm{rc}}$. We see that the modified net thermal flux curve for $\tau_{\rm{rc}}=0.1$ tends towards a constant value at low optical depths.  This behavior stems from (1) our removing the downwelling thermal flux boundary condition in the modified net flux expression, and (2) the general inability of low-opacity regions of the atmosphere to strongly emit or absorb radiative flux.  Finally note that, as $\tau$ approaches $\tau_0$, the modified net flux curves converge. This demonstrates that the dependence on $\tau_{\rm{rc}}$ is lost at large optical depths, which is expected as thermal fluxes are not sensitive to the radiative-convective boundary when this boundary is separated by many optical depths.  

\begin{figure}[p!]
\begin{center}
\includegraphics[width=4in]{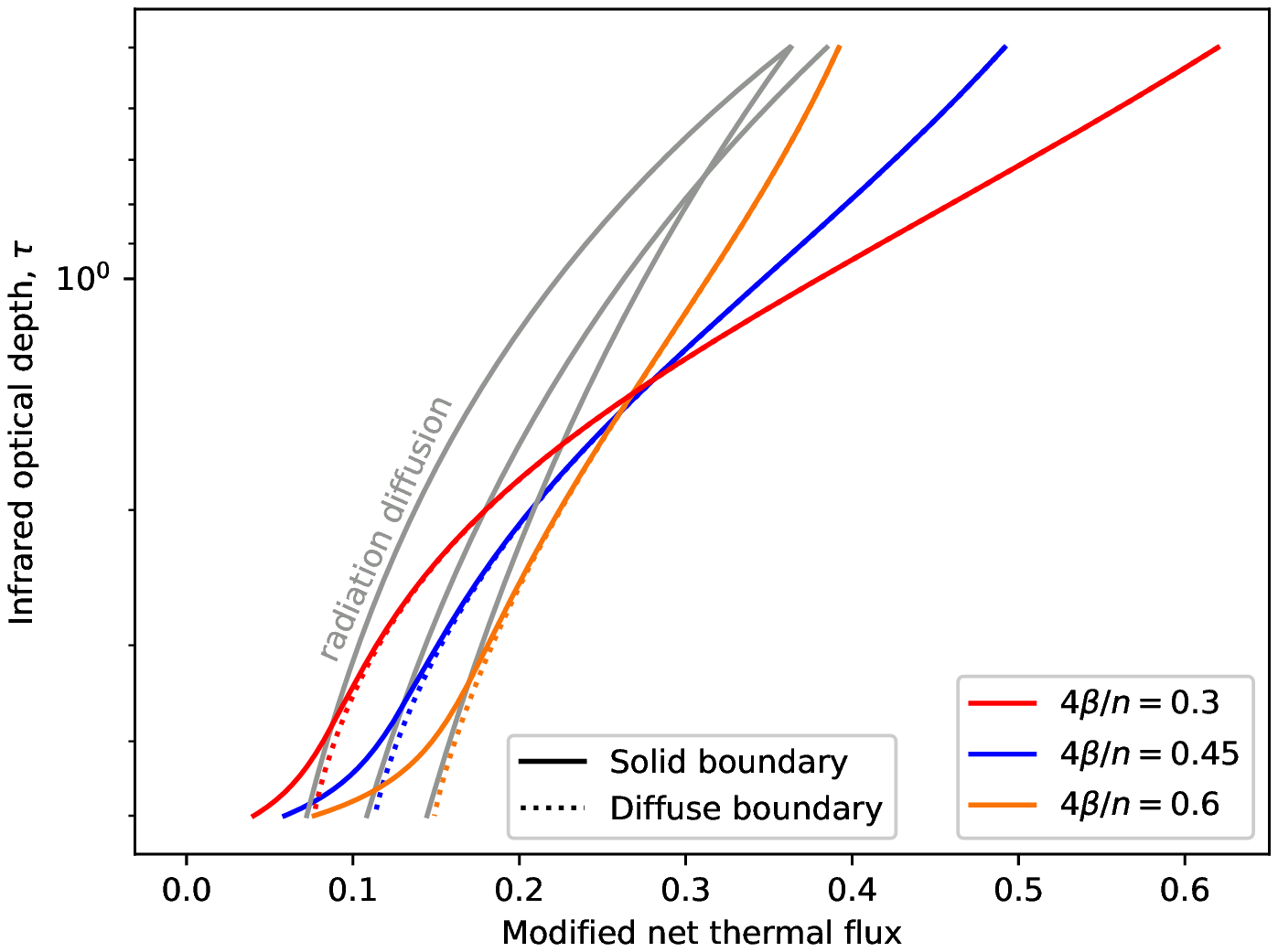}
\caption{
    \label{fig:varyw}
    Same as Figure~\ref{fig:varytau0}, except varying $4\beta/n$.  We fix $\tau_0=5$ and $\tau_{\rm{rc}}=0.5$.
    }
\end{center}
\end{figure}

Lastly, we explore variations in $4\beta/n$ in Figure~\ref{fig:varyw}.  We vary our nominal value ($4\beta/n=0.45$) to 
$4\beta/n = 0.3$ and $4\beta/n=0.6$, which are motivated by values seen in the tropospheres of Solar System worlds.  
As mentioned earlier, $4\beta/n$ controls the gradient in the Stefan-Boltzmann emittance in the convective 
portion of the atmosphere, and incorporates a modified dry adiabat (adjusted due to, e.g., latent heat release from 
condensation).  As can be seen in Figure~\ref{fig:varyw}, modified net thermal radiative flux models with 
different values of $4\beta/n$ approach their respective radiation diffusion limits at depth, and have distinct 
slopes (stemming from the different $T$-$\tau$ relationships) throughout the atmosphere.

%
\section{Comparative Planetology}
\label{sec:solsys}
%

\begin{figure}[p!]
\begin{center}
\includegraphics[width=4in]{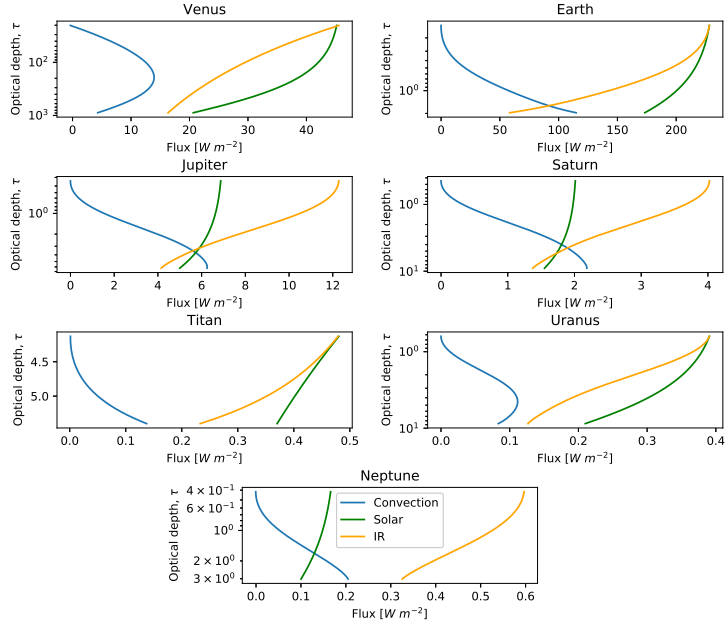}
\caption{
    \label{fig:subplots}
   Net solar, thermal, and convective fluxes in Solar System planetary atmospheres from our analytic treatment and adopting model parameters from Table~1.  For all plots, the convective flux is shown in blue, the net solar flux is shown in green, and the net infrared flux is shown in orange. All profiles are plotted for each world over their respective $\tau_{\rm{rc}}$--$\tau_0$ range, with the y-axis indicating $\tau$ (i.e., the grey thermal optical depth) on a log scale, with the exception of Titan which is plotted on a linear scale.  
    }
\end{center}
\end{figure}

We can use our analytic treatments to explore energy balances for a variety of worlds across the Solar System.  To accomplish this, we use Equation~\ref{eqn:fnetsol} to determine the net solar radiative flux, Equation~\ref{eqn:irfnet} to compute the net thermal radiative flux, and Equation~\ref{eqn:eqm} to find the convective heat flux.  Requisite model input parameters are taken from Table~1 of \citet{robinson&catling2014}, which were designed to reproduce the planetary average thermal structures of key Solar System worlds with thick atmospheres.  These parameters, as well as our parameters for Venus derived in Section~\ref{sec:valid}, are given in Table~\ref{tbl:params}.  In our analysis, we include Venus, Earth, Jupiter, Saturn, Titan, Uranus, and Neptune. These profiles of energy flux versus thermal optical depth are shown in Figure~\ref{fig:subplots}. We emphasize each world's troposphere, highlighting the respective region between $\tau_{\rm{rc}}$--$\tau_0$ for each planet. 

In our models and over the range of optical depths we investigate, Earth has the largest convective heat flux compared to any other world in our investigation.  These large convective heat fluxes for Earth result from (1) our planet's close proximity to the Sun paired with a moderately low albedo, leading to large amounts of absorbed solar radiation, and (2) Earth's relatively high atmospheric transparency to shortwave radiation, which implies that a significant amount of the absorbed solar radiation is deposited at the surface where thermal radiative transport is impeded by larger atmospheric thermal opacities. The role that this transparency plays can be seen in particular when comparing the convective heat flux profiles of Earth and Venus. As Venus has a thick, cloudy atmosphere, relatively little solar radiation is reaches the surface and deep atmosphere, resulting in a weakly convective atmosphere despite being closer to the Sun.

Flux profiles for Jupiter and Saturn show overall similar shapes.  Notice here that the net thermal radiative flux does not equal the net solar flux at the radiative-convective boundary, where the difference between these two quantities is simply the internal heat flux for each respective world.  Near the bottom of the depicted profiles, the net solar flux is rapidly being attenuated and the net thermal radiative flux is approaching the radiation diffusion limit.  Thus, at slightly larger optical depths than those shown here, the convective flux will simply approach the difference between the internal heat flux and the power-law radiation diffusion limit for the net thermal radiative flux.  Neptune shows similar behaviors to Jupiter and Saturn, except that the net thermal radiative flux is driven to relatively large values near (and above) the radiative-convective boundary by an internal heat flux that is roughly 60\% larger than the total absorbed solar flux.

Both Titan and Uranus demonstrate rather unique convective heat flux profiles.  For the former, and like Venus, overall weak convective fluxes are attributed to Titan's dense, opaque methane- and haze-rich atmosphere which absorbs a large amount of the incoming solar flux in the stratosphere, preventing solar energy from reaching the deep atmosphere and surface to drive convection.  In the case of Uranus, a turn-over in the convective heat flux is caused by differences in the shapes of the net solar versus net thermal fluxes (i.e., exponential drop-off versus power-law) and the near-zero internal heat flux \citep{pearletal1990}.

\begin{figure}[p!]
\begin{center}
\includegraphics[width=4in]{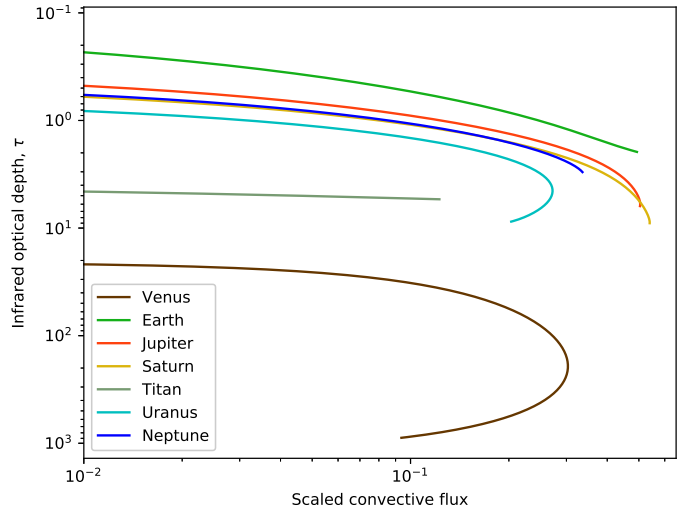}
\caption{
    \label{fig:allworlds}
   Scaled convective flux profiles for Solar System worlds, 
   obtained by dividing the model-derived convective flux profile by the input deep atmosphere energy flux budget (i.e., $F_2^{\odot} + F_{\rm{i}}$). Here, both the x-axis and y-axis are on a logarithmic scale.
    }
\end{center}
\end{figure}

We can gain a greater intuition for similarities between  convective profiles for Solar System worlds by scaling these profiles by the total tropospheric energy flux budget that must be carried by both convection and thermal radiative transport, $F^\odot_{2} + F_{\rm{i}}$.  As in \citet{robinson&catling2014}, $F^\odot_2$ is the net solar flux absorbed in the deep atmosphere and at the surface (if applicable).  Thus, $F_{\rm{c}}\left( \tau \right)$ scaled by $F^\odot_{2} + F_{\rm{i}}$ represents the fraction of the total deep atmosphere energy budget carried by convection.  These scaled profiles are shown in Figure~\ref{fig:allworlds}.  Aside from Titan and Venus (discussed below), the scaled convective flux profiles show an overall similar shape, where the convective flux begins to carry a significant portion (i.e., larger than 10\%) of the deep atmosphere energy budget when $D\tau$ reaches 0.7--2, which, intuitively, is where thermal radiative transport becomes inefficient due to increasingly opaque atmospheric conditions.

Regarding Titan and Venus, both have ``deep'' convective zones in the terminology of \citet{sagan1969}, where ``deep'' convective zones have $D\tau_{\rm{rc}}$ larger than unity and ``shallow'' convective zones have $D\tau_{\rm{rc}}$ smaller than unity.  As discussed in \citet{robinson&catling2012} (their Section~3.2), the division between ``deep'' and ``shallow'' convective zones is largely controlled by $\beta/n$ and the shortwave attenuation parameter(s), $k$.  The former parameter determines the critical lapse rate (in $T$-$\tau$ space) for the onset of convection.  For smaller stellar attenuation parameter(s), solar flux is deposited rapidly in the deep atmosphere or at the surface, leading to convective instability.  However, for larger stellar attenuation parameters (as in the case of Titan and Venus), the temperature profile is stabilized against convection by the absorption aloft of solar energy.

%
\section{Discussion}
%

Simple models of planetary thermal radiative transport and/or climate are only useful for understanding physical processes if such tools can be shown to reproduce observations or results from more-complex models.  Critically, our grey analytic expression of the net thermal radiative flux in a planetary atmosphere, Equation~\ref{eqn:irfnet}, provides strong reproductions of results from sophisticated spectrally-resolved models, as shown in Section~\ref{sec:valid}.  For Venus, our comparisons indicate that the deep atmosphere likely follows a $\tau \propto p$ relationship.  Also, our comparison to a cloudfree Jupiter shows that the grey assumption is an excellent approximation in the deep atmosphere, likely owing to the leading role played by H$_2$-H$_2$ and H$_2$-He pressure-induced absorption.  Thus, grey assumptions are likely to 
hold in the H$_2$- and He-rich deep atmospheres of gas giant exoplanets.

Our Earth comparison indicates that a $\tau \propto p^2$ relationship yields a good reproduction of the net thermal radiative flux profile computed by a line-by-line model.  Other authors have used the small scale height for water vapor in Earth's troposphere to argue for a steeper relationship between optical depth and pressure, with $\tau \propto p^4$ 
\citep{weaver&ramanathan1995,friersonetal2006}.  While it is certainly true that water is the dominant thermal opacity source in the deep atmosphere of Earth, it might be that cooler temperatures 1--2 pressure scale heights above Earth's surface shift the peak of the Planck function closer to the 15~$\mu$m CO$_2$ band, whose uniform mixing ratio through the atmosphere helps maintain $\tau \propto p^2$ in the deep atmosphere.

The uniform application of our analytic models to all Solar System worlds with thick atmospheres in Section~\ref{sec:solsys} enables exploration of commonalities (and differences) in convective flux profiles.  Owing to relatively strong shortwave attenuation, both Venus and Titan have ``shallow'' convective zones.  All other worlds --- Earth, Jupiter, Saturn, Uranus, and Neptune --- have ``deep'' convective zones.  These latter worlds also demonstrate a similar shape in their convective flux profiles (Figure~\ref{fig:allworlds}), once scaled by the deep atmosphere input energy budget (i.e., $F_2^{\odot}+F_{\rm{i}}$).  These scaled profiles show that convective transport rapidly increases when the grey thermal optical depth of the atmosphere reaches (roughly) unity.  Other authors have explored convective fluxes using grey thermal radiative transport principles, and we compare our approach to these previous studies in greater detail below.

\subsection{Ozawa and Ohmura}

\citet{ozawa&ohmura1997} adopted a grey thermal radiative transport model to explore the concept 
of ``maximum entropy production,'' wherein the steady-state radiative and convective energy fluxes 
are assumed to maximize the rate of entropy production.  In their work, \citet{ozawa&ohmura1997} 
varied the convective energy flux until entropy production was maximized, and thermal structures 
were derived from combining grey radiative transfer with an energy balance constraint (i.e., 
Equation~\ref{eqn:eqm}).  This approach does not require specification of a dry adiabatic lapse 
rate, and results in temperature profiles that are less-steep than their pure-radiative counterparts.  
Investigated cases all included an Earth-like net solar flux profile and total grey thermal optical 
depths (i.e., $\tau_0$) of between 1--5.

As is evident from Figure~2 in \citet{ozawa&ohmura1997}, equilibrium solutions for different 
values of $\tau_0$ are found to have distinct, non-constant $d\ln T/d\ln\tau$ values in the 
convective portion of an atmosphere, whereas this gradient is fixed as a constant ($\beta/n$) 
in our approach.  For \citet{ozawa&ohmura1997}, $d\ln T/d\ln\tau$ is typically 0.08--0.1 
beneath $\tau \sim 1$, falling to $<0.05$ in the less opaque regions of the atmosphere.  By 
comparison, our simulations (which assume an adiabatic, power-law temperature-pressure 
relationship) have $d\ln T/d\ln\tau$ equal to 0.17, 0.09, and 0.04 for $n$ equal to 1, 2, 
and 4, respectively.  Thus, our preferred Earth model (with $n=2$) has a markedly steeper 
$T-\tau$ relationship in the convective portion of the atmosphere than do solutions from 
\citet{ozawa&ohmura1997}.

Nevertheless, we can mimic solutions to equilibrium thermal structures in \citet{ozawa&ohmura1997} 
using the \citet{robinson&catling2012} model.  Instead of the typical approach, where equilibrium 
thermal structures are determined using energy balance and thermal structure continuity constraints 
to solve for $\tau_{\rm{rc}}$ and $\tau_0$ once $T_0$ (the surface temperature) and $\beta/n$ are 
specified, we can specify $T_0$ and $\tau_0$ \citep[both from Table~1 in][]{ozawa&ohmura1997} and 
find the values of $\tau_{\rm{rc}}$ and $\beta/n$ that yield an equilibrium thermal structure.  These 
models all have very ``shallow'' convective zones, with $\tau_{\rm{rc}}<0.1$ for $\tau_0$ ranging between 
1--5, whereas the Earth model given in \citet{robinson&catling2014} has $\tau_{\rm{rc}}=0.15$ and 
$\tau_0=2$.  Our models with fitted $T$-$\tau$ relationships also have smaller $d\ln T/d\ln\tau$, 
indicating less steep thermal structures.  This results in larger downwelling thermal radiative 
fluxes at the surface in our simulations, and, thus, larger convective convective fluxes at the 
planetary surface.

For a specific point of comparison, we investigate the \citet{ozawa&ohmura1997} case with 
$\tau_0 = 3$ and $T_0 = 289$, as this most closely matches Earth's mean surface temperature.  This 
particular case has a downwelling thermal radiative flux at the surface of 337~W~m$^{-2}$ and a 
convective flux at the surface of 89~W~m$^{-2}$ \citep[see Table~1 in][]{ozawa&ohmura1997}.  By 
comparison, when we adopt a model that fits for $\tau_{\rm{rc}}$ and $\beta/n$ while using the 
same values of $\tau_0$ and $T_0$, we find a downwelling 
thermal radiative flux at the surface of 370~W~m$^{-2}$ and a surface convective heat flux of 
120~W~m$^{-2}$.  Estimates of Earth's surface energy budget typically find a downwelling thermal 
radiative flux at the surface of roughly 330--340~W~m$^{-2}$ and a surface convective heat flux (in both 
dry and moist processes) of 100--110~W~m$^{-2}$ \citep{trenberthetal2009,loebetal2009}.  Thus our 
``fitted $\beta/n$'' approach overestimates the downwelling thermal radiative flux at the surface by 
about 10\%, resulting in a similar overestimate in the convective heat flux.  

Our more realistic Earth model (adopted in Section~\ref{sec:solsys}) has 330~W~m$^{-2}$ of downwelling 
thermal radiative flux and 110~W~m$^{-2}$ of convective heat flux at the surface, which are 
quite close to the global average estimates and comparable to the \citet{ozawa&ohmura1997} case.  
A key difference is that our adopted $d\ln T/d\ln\tau$ is motivated by the physics of (modified) dry 
adiabats and pressure-broadened opacities rather than maximizing the rate of entropy production.  Energy 
flux agreements between our approach and the maximum entropy generation technique indicate that our 
analytic models could be used to explore physical underpinnings in maximum entropy generation models.

\subsection{Lorenz and McKay}

\citet{lorenz&mckay2003} used expressions for grey thermal radiative equilibrium and realistic 
data and models to propose an empirical expression for the convective flux at a planetary surface 
of the form,
\begin{equation}
    F_{\rm{c}} = F_{\rm{net}}^{\odot}\left( \tau_0 \right) \frac{\tau_0}{C + D\tau_0} \ ,
\end{equation}
where $C$ and $D$ are free parameters (the latter is distinct from the diffusivity factor 
used above).  Typical values for both $C$ and $D$ were in the range 1--2.  Convective fluxes in 
giant planets is only briefly discussed by \citet{lorenz&mckay2003}, who note that the deep 
atmospheres of gas giants will have large $\tau_0$ and $F_{\rm{c}} \approx F_{\rm{i}}$, indicating 
that replacing $F_{\rm{net}}^{\odot}\left( \tau_0 \right)$ with $F_{\rm{i}}$ in the empirical 
expression and adopting $D=1$ would fit the deep atmospheres of giants.

\begin{figure}[p!]
\begin{center}
\includegraphics[width=4in]{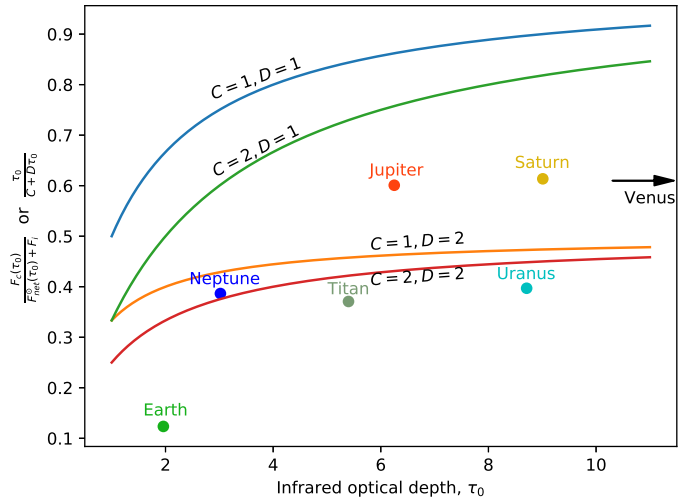}
\caption{
    \label{fig:l&m03}
    \label{lastfig}
   Variants of the heuristic convective flux expression 
   given by \citet{lorenz&mckay2003} compared to the Solar 
   System worlds explored with our analytic model.
    }
\end{center}
\end{figure}

We investigate the applicability of the empirical expression from \citet{lorenz&mckay2003} in 
Figure~\ref{fig:l&m03}.  Inclusion of the gas and ice giants is not straightforward, as these worlds 
lack a ``surface'' where the empirical expression is designed to apply.  Nevertheless, as 
the convective flux in the tropospheres of these worlds must carry some fraction of the 
combined net solar and internal heat fluxes, we opt to plot 
$F_{\rm{c}}\left(\tau_0\right)/\left[F_{\rm{net}}^{\odot}\left( \tau_0 \right) + F_{\rm{i}}\right]$ 
versus $\tau_0$ for the worlds in Section~\ref{sec:solsys}.  For terrestrial worlds, the optical depth 
is referenced at the surface, while for gaseous worlds we use the 1~bar pressure level.  Of course, 
adopting the 1 bar pressure level for gaseous worlds is arbitrary, and using other pressure levels 
would result in values of 
$F_{\rm{c}}\left(\tau_0\right)/\left[F_{\rm{net}}^{\odot}\left( \tau_0 \right) + F_{\rm{i}}\right]$ 
that span from 0 at the radiative-convective boundary and unity at great depths.  Focusing only on solid-surface planets, the shown variations of the \citet{lorenz&mckay2003} expression do not offer perfect fits, although the performance of this expression is surprising given its two-parameter simplicity.

%
\section{Conclusions}
%

Net thermal and convective energy fluxes are critically important to determining atmospheric thermal structure, especially in one-dimensional (vertical) planetary climate models.  We derive a simple expression for the net downwelling thermal radiative flux in a planetary troposphere where the relationships between temperature, pressure, and optical depth are all expressed as power-laws.  When combined with previous results, our new treatment yields an analytic expression for the net thermal radiative flux in a convective planetary troposphere.  For appropriate, physically-based input parameters, our analytic net thermal radiative flux expression reproduces results from more-sophisticated, spectrally-resolved models applied to Earth, Venus, and a cloudfree Jupiter.  Application of our model across the Solar System demonstrates common shapes and scalings in the convective flux profiles of Earth, Jupiter, Saturn, Uranus, and Neptune.  Further applications of our model sheds new light on both ``maximum entropy production'' principles as well as other simple treatments of convection.  Simple models remain an excellent tool for inter-comparing processes in planetary atmospheres both within and beyond the Solar System.

%
\ack
TDR is supported by an award from the NASA Exoplanets Research Program (\#80NSSC18K0349) and by the NASA Astrobiology Institute's Virtual Planetary Laboratory under Cooperative Agreement \#NNA13AA93A.  JPT is supported by the National Science Foundation's Research Experience for Undergraduates program through the CAMPARE program directed out of California State Polytechnic University, Pomona.
%

\label{lastpage}





\clearpage	

\begin{table}
\begin{center}
\textbf{Radiative-Convective Model Parameters for Solar System Worlds}
\begin{tabular}{llllllll}
\hline 
\hline
{\bf World}&{\bf Venus}&{\bf Earth}&{\bf Jupiter}&{\bf Saturn}&{\bf Titan}&{\bf Uranus}&{\bf Neptune} \\
\hline
 $p_0$ (bar) & 92.1 & 1 & 1 & 1 & 1.5 & 1 & 1 \\
 $T_0$ (K)  & 730 & 288 & 166 & 135 & 94 & 76 & 72 \\
 $a$  & 0.8 & 0.6 & 0.85 & 0.94 & 0.77 & 0.83 & 0.87 \\
 $\gamma$  & 1.3 & 1.4 & 1.4 & 1.4 & 1.4 & 1.4 & 1.4 \\
 $n$  & 1 & 2 & 2 & 2 & 2 & 2 & 2 \\
 $F_1^\odot$ (W~m$^{-2}$)  & 110 & 7 & 1.3 & 0.41 & 1.49 & 0.24 & 0.09 \\
 $F_2^\odot$ (W~m$^{-2}$)  & 46 & 233 & 7.0 & 2.04 & 1.12 & 0.41 & 0.18 \\
 $F_{\rm{i}}^\odot$ (W~m$^{-2}$)  & 0 & 0 & 5.4 & 2.01 & 0 & $\sim$ 0 & 0.43 \\
 $k_1$   & 0.518 & 90 & 90 & 180 & 120 & 220 & 580 \\
 $k_2$   & $9\times 10^{-4}$ & 0.16 & 0.05 & 0.03 & 0.2 & 0.08 & 0.2 \\
 $\tau_{\rm{rc}}$ & 20 & 0.15 & 0.34 & 0.44 & 4.4 & 0.63 & 0.42 \\
 $\tau_0$ & 990 & 1.9 & 6.3 & 9.0 & 5.6 & 8.7 & 3.0 \\
\hline
\end{tabular}
\caption[Radiative-Convective Model Parameters]	
	{
	\label{tbl:params}	
	\label{lasttable}
	Model parameters used in our analytic expressions.  Venus parameters come from fits in Section~\ref{sec:valid}.  Parameters for all other worlds come from \citet{robinson&catling2014} with an updated diffuse lower boundary condition for Jupiter, Saturn, Uranus, and Neptune.
	}
\end{center}
\end{table}

\clearpage


\end{document}